\documentclass[11pt]{article}
\usepackage[pctex32]{graphics}

\def\ben{\begin{equation}}
\def\een{\end{equation}}

\def\nn{\nonumber} \def\bd{\begin{document}} \def\ed{\end{document}}
\def\ds{\documentstyle} \let\fr=\frac \let\bl=\bigl \let\br=\bigr
\let\Br=\Bigr \let\Bl=\Bigl
\let\bm=\bibitem
\let\na=\nabla
\let\pa=\partial \let\ov=\overline
\newcommand{\be}{\begin{equation}}
\newcommand{\ee}{\end{equation}}
\def\ba{\begin{array}}
\def\ea{\end{array}}
\def\ft#1#2{{\textstyle{\frac{\scriptstyle #1}{\scriptstyle #2} } }}
\def\fft#1#2{{\frac{#1}{#2}}}
\def\del{\partial}
\def\vp{\varphi}
\def\sst#1{{\scriptscriptstyle #1}}
\def\oneone{\rlap 1\mkern4mu{\rm l}}
\def\td{\tilde}
\def\wtd{\widetilde}
\def\ie{{\it i.e.\ }}
\def\dalemb#1#2{{\vbox{\hrule height .#2pt
        \hbox{\vrule width.#2pt height#1pt \kern#1pt
                \vrule width.#2pt}
        \hrule height.#2pt}}}
\def\square{\mathord{\dalemb{6.8}{7}\hbox{\hskip1pt}}}
\newcommand{\ho}[1]{$\, ^{#1}$}
\newcommand{\hoch}[1]{$\, ^{#1}$}
\newcommand{\bea}{\setlength\arraycolsep{2pt} \begin{eqnarray}}
\newcommand{\eea}{\end{eqnarray}}
\newcommand{\ra}{\rightarrow}
\newcommand{\lra}{\longrightarrow}
\newcommand{\Lra}{\Leftrightarrow}
\newcommand{\bp}{\tilde \beta^\prime}
\newcommand{\tr}{{\rm tr} }
\newcommand{\Tr}{{\rm Tr} }
\def\0{{\sst{(0)}}}
\def\1{{\sst{(1)}}}
\def\2{{\sst{(2)}}}
\def\3{{\sst{(3)}}}
\def\4{{\sst{(4)}}}
\def\5{{\sst{(5)}}}
\def\6{{\sst{(6)}}}
\def\7{{\sst{(7)}}}
\def\8{{\sst{(8)}}}
\def\m{{\sst{(m)}}}
\def\n{{\sst{(n)}}}
\def\cA{{{\cal A}}}
\def\cB{{{\cal B}}}
\def\cF{{{\cal F}}}
\def\cG{{{\cal G}}}
\def\cH{{{\cal H}}}
\def\tV{\widetilde V}
\def\tW{\widetilde W}
\def\tH{\widetilde H}
\def\tE{\widetilde E}
\def\tF{\widetilde F}
\def\tA{\widetilde A}
\def\im{{{\rm i}}}
\def\tY{{{\wtd Y}}}
\def\ep{{\epsilon}}
\def\vep{{\varepsilon}}
\def\bD{{{\bar D}}}
\def\R{{{\mathbb R}}}
\def\C{{{\mathbb C}}}
\def\H{{{\mathbb H}}}
\def\CP{{{\mathbb C}{\mathbb P}}}
\def\RP{{{\mathbb R}{\mathbb P}}}
\def\Z{{{\mathbb Z}}}
\def\bA{{{\mathbb A}}}
\def\bB{{{\mathbb B}}}
\def\bC{{{\mathbb C}}}
\def\bD{{{\mathbb D}}}
\def\bE{{{\mathbb E}}}
\def\bZ{{{\mathbb Z}}}
\def\Re{{{\frak{Re}}}}
\def\Im{{{\frak{Im}}}}
\def\cosec{{\,\hbox{cosec}\,}}
\def\Gm{{\Gamma_{\!\! -}}}
\def\Gp{{\Gamma_{\!\! +}}}
\def\stan{{standard }}
\def\nonstan{{supernumerary }}
\def\p{{\partial}}
\def\kdel#1{{\fft{\del}{\del#1}}}

\def\bog{{Bogomolny }}
\def\om{{\omega}}

\newcommand{\nnr}{\nonumber \\}
\newcommand{\pd}{\partial}
\newcommand{\ud}{\textrm{d}}
\newcommand{\dTH}{T^{\prime \, 0}_\textrm{H}}
\newcommand{\dOi}{\Omega^{\prime \, 0}_i}
\newcommand{\bx}{{\bf x}}

\begin{document}

\vspace{5mm}
\begin{center}
{\Large \bf Generalized uncertainty principle and
Ho\v{r}ava-Lifshitz gravity } \vspace{12mm}

{\large   Yun Soo Myung \footnote{e-mail
 address: ysmyung@inje.ac.kr}}
 \\
\vspace{10mm} {\em Institute of Basic Science and School of
Computer Aided Science \\ Inje University, Gimhae 621-749, Korea}
\end{center}

\begin{center}

\underline{Abstract}
\end{center}

  We explore a connection between generalized uncertainty
  principle (GUP) and modified Ho\v{r}ava-Lifshitz (HL) gravity.
The GUP density function may  be replaced by the cutoff function
for the renormalization group of modified Ho\v{r}ava-Lifshitz
gravity. We find the GUP-corrected graviton propagators and
compare these with tensor propagators in the HL gravity. Two are
qualitatively similar, but  the $p^5$-term arisen from Cotton
tensor is missed in the GUP-corrected graviton propagator.

\vspace{15pt}

\thispagestyle{empty}





\newpage
\section{Introduction}
Recently Ho\v{r}ava has proposed a renormalizable theory of
gravity at a Lifshitz point~\cite{ho1},  which  may be regarded as
a UV complete candidate for general relativity. At short distances
the theory of $z=3$ Ho\v{r}ava-Lifshitz (HL) gravity describes
interacting nonrelativistic gravitons and is supposed to be power
counting renormalizable in (1+3) dimensions. Recently, the HL
gravity theory has been intensively investigated
in~\cite{ho2,Vi,ho3,VW,Kl1,Ni,Na,Iz,Vo,SVW1,CH,CHZ,Nis,OR,KLM1,Ko,LP,CNPS,SVW2,Ca,Sa,My,GKS,BPS,Kob,BS},
 its cosmological applications in
~\cite{cos1,cos2}, and its black hole solutions in
~\cite{LMP,KS,CCO1}.

It seems that the GUP effect on the Schwarzschild black hole is
related to black holes in the deformed Ho\v{r}ava-Lifshitz
gravity~\cite{Myung}.  We could not confirm a solid connection
between the GUP and the black hole of modified Ho\v{r}ava-Lifshitz
gravity, although we have obtained partial connections between
them.

However,  it was known that the generalized uncertainty principle
provides naturally a UV cutoff to the local quantum field theory
as gravity effects~\cite{CMOK,KLM2}.

It is known that the UV-propagator for tensor modes $t_{ij}$ take
a complicated form Eq. (\ref{tenprp}) including upto $p^6$-term
from the Cotton bilinear term  $C_{ij}C_{ij}$. At low energies,
the UV-propagator may reduce to a conventional IR-propagator as
$G_{\rm IR}(\omega,\vec{p})=1/(\omega^2-c^2\vec{p}^2)$ for $z=1$
HL gravity. It is very important to understand why the
UV-propagator takes a complicated form in the non-relativistic
gravity theory.

In this work, we investigate a connection between GUP and modified
Ho\v{r}ava-Lifshitz  gravity. The GUP density function may  be
replaced by a cutoff function for the renormalization group study
of modified Ho\v{r}ava-Lifshitz gravity. We find GUP-corrected
graviton propagators and compare these with UV-tensor propagators
in the HL gravity. Two are similar, but the $p^5$-term arisen from
Cotton tensor is missed in the GUP-corrected graviton propagator.
This shows that  a power-counting renormalizable theory of the HL
gravity is closely related to the GUP.

 \section{HL gravity}
Introducing the ADM formalism where the metric is parameterized
\cite{adm}
\be ds_{ADM}^2= - N^2  dt^2 + g_{ij} \Big(dx^i - N^i dt\Big)
\Big(dx^j - N^j dt\Big)\,, \ee
the Einstein-Hilbert action can be expressed as
\be \label{Eins} S_{EH} = \fft{1}{16\pi G} \int d^4x \sqrt{g} N
\Big[K_{ij} K^{ij} - K^2 + R - 2\Lambda\Big], \ee
where $G$ is Newton's constant and extrinsic curvature $K_{ij}$
takes the form
\be K_{ij} = \fft{1}{2N} \Big(\dot g_{ij} - \nabla_i N_j -
\nabla_j N_i\Big)\,. \ee
Here, a dot denotes a derivative with respect to $t$. An action of
the non-relativistic renormalizable gravitational theory  is given
by~\cite{ho1} \be S_{HL}=\int dtd^3x \Big[{\cal L}_K + {\cal
L}_V\Big],  \label{action1} \ee where the kinetic terms are given
by \be {\cal L}_K =\frac{2}{\kappa^2}\sqrt{g} N K_{ij}{\cal
G}^{ijkl}K_{kl}= \frac{2}{\kappa^2}\sqrt{g}
N\Big(K_{ij}K^{ij}-\lambda K^2\Big), \ee with the DeWitt metric
 \be {\cal G}^{ijkl}=
\frac{1}{2}\Big(g^{ik}g^{jl}-g^{il}g^{jk}\Big)-\lambda
g^{ij}g^{kl} \ee
 and its inverse metric
 \be {\cal
G}_{ijkl}=\frac{1}{2}\Big(g_{ik}g_{jl}-g_{il}g_{jk}\Big)-\frac{\lambda}{3\lambda-1}g_{ij}g_{kl}.\ee

The potential terms is determined by the detailed balance
condition (DBC) as \bea {\cal L}_V=-\frac{\kappa^2}{2}\sqrt{g}N
E^{ij}{\cal G}_{ijkl}E^{kl}&=&
\sqrt{g}N\Bigg\{\frac{\kappa^2\mu^2}{8(1-3\lambda)}\Big(\frac{1-4\lambda}{4}R^2
+\Lambda_WR-3\Lambda_W^2\Big)\nn \\
 &-&\frac{\kappa^2}{2w^4} \left(C_{ij}
-\frac{\mu w^2}{2}R_{ij}\right) \left(C^{ij} -\frac{\mu
w^2}{2}R^{ij}\right) \Bigg\}.\label{action2} \eea Here the $E$
tensor is defined by \be E^{ij}=\frac{1}{w^2} C^{ij}-\frac{\mu}{2}
\Big(R^{ij}-\frac{R}{2} g^{ij}+\Lambda_Wg^{ij}\Big) \ee with the
Cotton tensor $C_{ij}$ \be
C^{ij}=\frac{\epsilon^{ik\ell}}{\sqrt{g}}\nabla_k\left(R^j{}_\ell
-\frac14R\delta_\ell^j\right).\label{def.K.C} \ee  Explicitly,
$E_{ij}$ could be derived  from the Euclidean topologically
massive gravity \be E^{ij}=\frac{1}{\sqrt{g}} \frac{\delta
W_{TMG}}{\delta g_{ij}} \ee with \be W_{TMG}=\frac{1}{w^2} \int
d^3 x \epsilon^{ikl}\Big(\Gamma^m_{il}\partial_j
\Gamma^l_{km}+\frac{2}{3} \Gamma^n_{il} \Gamma^l_{jm}
\Gamma^m_{kn} \Big)- \mu \int d^3x \sqrt{g}(R-2\Lambda_W), \ee
where $\epsilon^{ikl}$ is a tensor density with
$\epsilon^{123}=1$.

In the IR limit,  comparing ${\cal L}_0$ with Eq.(\ref{Eins}) of
general relativity, the speed of light, Newton's constant and the
cosmological constant are given by
\be c=\fft{\kappa^2\mu}{4}
\sqrt{\fft{\Lambda_W}{1-3\lambda}}\,,\qquad
G=\fft{\kappa^2}{32\pi\,c}\,,\qquad \Lambda_{\rm cc}=\ft32
\Lambda_W\,.\label{cg} \ee The equations of motion were derived in
\cite{cos1} and \cite{LMP}. We would like to mention that the IR
vacuum of this theory is anti-de Sitter (AdS$_4$) spacetimes.
Hence, it is interesting to take a limit of the theory, which may
lead to  a Minkowski vacuum in the IR sector. To this end, one may
deform the theory by introducing ``$\mu^4R$" $(\tilde{{\cal
L}}_V={\cal L}_V+\sqrt{g}N \mu^4R)$ and then, take the $\Lambda_W
\to 0$ limit~\cite{KS}. This does not alter the UV properties of
the theory, while it changes the IR properties. That is, there
exists a Minkowski vacuum, instead of an AdS vacuum. In the IR
limit, the speed of light and  Newton's constant are given by
\be c^2=\fft{\kappa^2\mu^4}{2},~ G=\fft{\kappa^2}{32\pi\,c},
~\lambda=1.\label{kh} \ee

\section{GUP}
A meaningful prediction of various theories of quantum gravity
(string theory) and black holes is the presence of a minimum
measurable length or a maximum observable momentum. This has
provided the generalized uncertainty principle which modifies
commutation relations between position coordinates and momenta.
Also  the black hole solution of modified HL gravity reminds us
the Schwarzschild black hole modified with GUP~\cite{CMOK}. Hence,
we make
 a  close connection between GUP and
Ho\v{r}ava-Lifshitz gravity.  A commutation relation of
         \begin{equation} \label{crgup}
         [\vec{x}, \vec{p}]=i\hbar(1+\beta^2\vec{p}^2)
         \end{equation}
         leads to the  generalized uncertainty relation
 \be \label{1eq2} \Delta x \Delta p \ge \frac{\hbar}{2}
\Bigg[1+\alpha^2 l_p^2 \frac{(\Delta p)^2}{\hbar^2}\Bigg] \ee with
$l_p=\sqrt{G \hbar/c^3}$ the Planck length. Here  a parameter
$\alpha=\hbar\sqrt{\beta}/l_p$ is introduced to indicate the GUP
effect. The Planck mass is given by $m_p=\sqrt{\hbar c/G}$. The
above implies a lower bound on the length scale \be \label{2eq2}
\Delta x \ge (\Delta x)_{\rm mim} \approx \hbar \sqrt{\beta}=
\alpha l_p, \ee
 which means that the Planck length plays
the role of a fundamental scale. On the other hand, Eq.
(\ref{1eq2}) implies  the upper bound on  the momentum as \be
\label{2eq2-1} \Delta p \le (\Delta p)_{\rm max} \approx
\frac{1}{\sqrt{\beta}}= \frac{m_p c}{\alpha}. \ee
\begin{figure}[t!]
   \centering
   \includegraphics{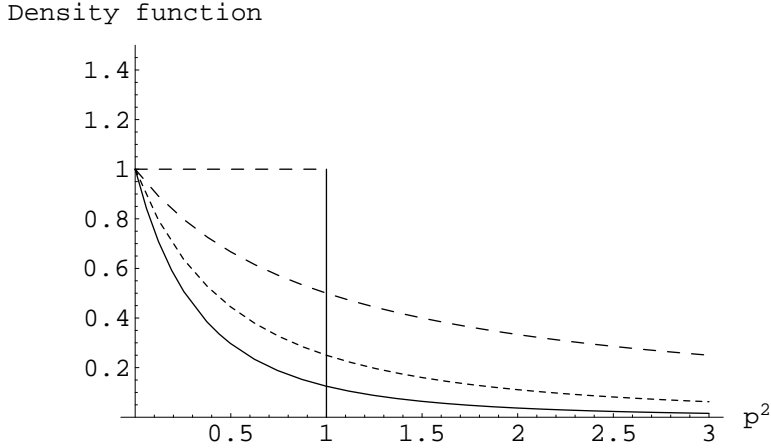}
\caption{Density functions for regularization of  a LQFT as
function of $p^2$ with $\beta=1$. The dashed line denotes
``arbitrary uniform density function" as the UV cutoff
$1/\sqrt{\beta}=1$ for $ p^2 \in[0,1]$ required by hand, while
three curves represent the GUP density function in
Eq.(\ref{ddgup}) for $D=1,2,$ and 3 from top to bottom. These
curves cut effectively off the integral beyond $p=1/\sqrt{\beta
}$.} \label{fig.1}
\end{figure}
Importantly,  it was known that the generalized uncertainty
principle provides naturally a UV cutoff to the local quantum
field theory (LQFT) as gravity effects~\cite{CMOK,KLM2}. The GUP
relation of Eq.(\ref{crgup}) has an effect on the density of
states in $D$-dimensional momentum space as
\begin{equation} \label{dgup}
d^D\vec{p}~{\cal D}_D(\beta\vec{p}^2),
\end{equation}
where a density function (weight factor) ${\cal
D}_D(\beta\vec{p}^2)$ is defined by \be \label{ddgup}{\cal
D}_D(\beta\vec{p}^2)=\frac{1}{(1+\beta \vec{p}^2)^D}.\ee As is
depicted in Fig. 1, this function cuts  effectively off the
integral beyond $p=1/\sqrt{\beta}$. Due to strong suppression of
density of states at high momenta, a relevant quantity will be
rendered finite with $1/\sqrt{\beta}$ acting effectively as a UV
cutoff.  We wish to mention that this function may be related to
the Cotton-term of $C_{ij}C^{ij}$ in Eq. (\ref{action2}) because
the latter contains a sixth order derivative. We note that the
arbitrary uniform density function is introduced by hand and thus,
the physics beyond the cutoff ($\vec{p}^2>1/\sqrt{\beta}=1$) never
contributes to a relevant quantity.

The right-hand side of Eq. (\ref{crgup}) includes a
$\vec{p}$-dependent term  and thus affect the cell size in phase
space as ``being $\vec{p}$-dependent". Making use of the Liouville
theorem, one could show that  the invariant weighted-phase space
volume under time evolution is given by~\cite{CMOK}
\begin{equation} \label{sgup}
\frac{d^D\vec{x}d^D\vec{p}}{(1+\beta \vec{p}^2)^D},
\end{equation}
where the classical commutation relations corresponding to the
quantum commutation relation of Eq. (\ref{crgup}) are given via
$[A,B]/i\hbar \to \{A,B\}$ by \be \{x_i, p_j\}=(1+\beta
p^2)\delta_{ij},~~ \{p_i,p_j\}=0, ~~
\{x_i,x_j\}=2\beta(p_ix_j-p_jx_i).\ee
 Actually, $1/\sqrt{\beta}$
plays the role of a  UV cutoff  $\Lambda$ of the momentum
integration as~\cite{KLM2} \be \frac{1}{\sqrt{\beta}}\to
\Lambda.\ee As a concrete example, by assuming that the zero-point
energy of each oscillator is of $\hbar \omega /2=\hbar
\sqrt{\vec{p}^2+m^2}/2$, the cosmological constant is calculated
to be \be \Lambda_{\rm CC}(m)=\int \frac{d^3\vec{p}}{(1+\beta
\vec{p}^2)^3}\Big[\frac{\sqrt{\vec{p}^2+m^2}}{2}\Big]=2\pi
\int_0^\infty \frac{p^2dp}{(1+\beta
p^2)^3}\sqrt{p^2+m^2}=\frac{\pi}{2\beta^2}f(\beta m^2), \ee with
$f(0)=1$ and $\hbar=1$. Then,  one obtains the cosmological
constant for the massless case\be \Lambda_{\rm
cc}(0)=\frac{\pi}{2\beta^2} \to \frac{\pi}{2}\Lambda^4. \ee
Finally, the GUP commutation relation in Eq. (\ref{crgup}) can be
extended into~\cite{Nou}

         \begin{equation}
         [\vec{x}, \vec{p}]=i\hbar e^{\beta^2\vec{p}^2},
         \end{equation}
which  includes all order corrections to the Heisenberg
uncertainty principle. In this case,  the density function   is
given by an exponential function ~\cite{KP}
\begin{equation} \label{allgup}
{\cal D}^{\rm
all}_{D}(\beta\vec{p}^2)=\frac{1}{e^{\beta^2\vec{p}^2}}.
\end{equation}

\section{Cutoff function for a relativistic theory}

It is well known that  even the simplest local quantum field
theories (LQFT) are useless because the answer to any loop
calculation is infinite. A standard example is the 1-loop
correction to the mass in scalar
$\tilde{\lambda}\phi^3$~\cite{sho} \bea \Delta
m^2&=&\frac{\tilde{\lambda}^2}{2}\int
\frac{d^4p}{(2\pi)^4}\frac{1}{(p^2+m^2)((p+q)^2+m^2)}\\
&=& {\rm finite}+ {\cal C} \int^{\infty}\frac{dp}{p}=\infty \eea
with $p^2=p_\mu p^\mu$ and a constant ${\cal C}$. The reason why
we have this meaningless result is clear  because of integrating
all the way to infinity in momentum space.  One way to avoid
infinity is to introduce a UV  cutoff $\Lambda$.  However, we run
into trouble with LQFT, because  a regularized theory is no longer
unitary since we ``arbitrarily" removed part of the phase space
(by hand) to which there was associated a non-zero amplitude. In
order to find an appropriate situation, we wish to probe  the
system at some energy scale $\Lambda_R$ namely, incoming momenta
in Feynman graphs obey $p\leq\Lambda_R$, while keeping $\Lambda\gg
\Lambda_R$.  If we can make all physical observables at
$\Lambda_R$ independent of $\Lambda$, then we can safely take
$\Lambda\rightarrow\infty.$
\begin{figure}[t!]
   \centering
   \includegraphics{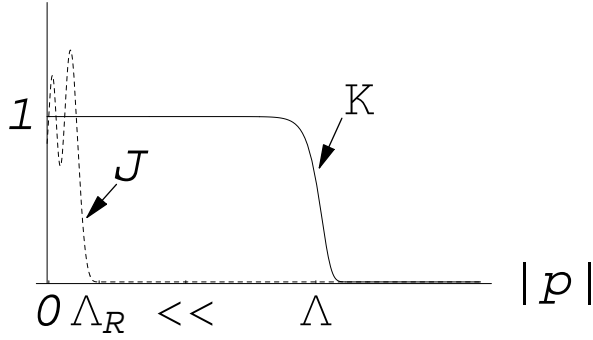}
\caption{The linear source term is constrained to $J(p)=0$ for
$p>\Lambda_R$ so as to only excite Green functions with low
energy. The quadratic term contains a cutoff function
$\mathcal{K}(\frac{p^2}{\Lambda^2})$ with the property that
$\mathcal{K}=1$ for $p<\Lambda$ and then falls off smoothly to
zero for $p\geq\Lambda$.} \label{fig.2}
\end{figure}
 It is convenient to parameterize the energy scale using
an RG ``time'' parameter flowing towards lower and lower energies
\begin{equation}\label{rgt}
\Lambda(t)=\Lambda(0)e^{-t}.
\end{equation} Let us  demand that  changing the cutoff $\Lambda$
leaves the partition function invariant as
\begin{equation}\label{dtzisz}
\partial_tZ[J]=0.
\end{equation}
Then we could define the partition function of a LQFT  by
\begin{equation}
Z[J]=\frac{I[J]}{I[0]}
\end{equation}
where \be
I[J]=\int[d\phi]e^{-\bigl(\mathcal{S}_0+\mathcal{S}_I+\mathcal{S}_J\bigr)}.
\ee The action is composed of linear source term,  quadratic
kinetic term and polynomial interaction term  as \bea
\label{source1}\mathcal{S}_J&=&\int\frac{d^4p}{(2\pi)^4}J(p,\Lambda_R)\phi(-p),\\
\label{source2} \mathcal{S}_0&=&\frac{1}{2}\int\frac{d^4p}{(2\pi)^4}\phi(p)\phi(-p)\frac{\tilde{\Delta}(p^2)}{\mathcal{K}(\frac{p^2}{\Lambda^2})},\\
\label{source3}\mathcal{S}_I&=&\sum_{n=3}^{\infty}\int\frac{d^4p_1}{(2\pi)^4}\cdots\frac{d^4p_n}{(2\pi)^4}\delta^4\bigl(\sum
p_i\bigr) g_n(p_1,\cdots,p_n;\Lambda)\phi(p_1)\cdots\phi(p_n),
\eea where $\tilde{\Delta}(p^2)$ is the inverse propagator in
four-momentum space.
 That is, $\tilde{\Delta}(p^2)=p^2+m^2$ for a massive scalar field. Here
  we include a source function $J(p)$ and a smooth cutoff function $\mathcal{K}(p^2/\Lambda^2)$ with
  specific properties as was described  in Fig. 2.
From Eq. (\ref{source2}), a relativistic propagator could be
derived to take the form \be \tilde{\Delta}(p^2)^{-1} \times
\mathcal{K}(\frac{p^2}{\Lambda^2}). \ee We mention two important
properties: $\partial_t\mathcal{K}\cdot J=0$ because they have
disjoint support  and  $\partial_t J(p)=0$ because $J$ depends
only on $\Lambda_R$. Finally, using these properties, one arrives
at
\begin{equation}\label{dtsii}
\partial_t\bigl(e^{-S_I}\bigr)=-\frac{1}{2}\int\frac{d^4p}{(2\pi)^4}
\frac{\partial_t\mathcal{K}}{\tilde{\Delta}(p^2)}\frac{\delta}{\delta\phi(p)}\frac{\delta}{\delta\phi(-p)}e^{-\mathcal{S}_I}.
\end{equation}
Eq. (\ref{dtsii}) describes the infinitesimal change of the
interaction Lagrangian upon changing the UV cutoff $\Lambda.$ This
dependence of the coupling constants on the cutoff $\Lambda$  is
called the ``RG flow". A procedure of decreasing the cutoff on
$|p|$ infinitesimally from $\Lambda$ to $\Lambda-\delta\Lambda$ is
called  integrating out a momentum shell. We note  that  no
infinities are encountered here because all the momentum integrals
are done in an infinitesimally finite range.

Finally, we propose that  the cutoff function
$\mathcal{K}(p^2/\Lambda^2)$ for a relativistic theory can be
replaced by the density function ${\cal D}_D(\beta\vec{p}^2)$ for
a non-relativistic gravity theory: \be \mathcal{K}(p^2/\Lambda^2)
\to {\cal D}_D(\beta\vec{p}^2). \ee This is quite reasonable
because two functions play the similar role in suppressing high
momenta (UV region).

\section{ Propagators of HL gravity}
We wish to consider perturbations  of the metric around Minkowski
spacetimes, which is  a solution to the $z=3$ HL gravity
(\ref{action1}) \be
g_{ij}=\delta_{ij}+wh_{ij},~~N=1+wn,~~N_i=wn_i. \ee
 In order to have tensor propagator, it is convenient to use the
cosmological decomposition in terms of scalar, vector, and tensor
modes under spatial rotations $SO(3)$~\cite{MFB}
 \bea \label{pert}
 n &=&-\frac{1}{2}A,\nn \\
 n_i&=&\partial_iB+V_i,\label{decom2} \\
 h_{ij}&=&\psi\delta_{ij}+\partial_i\partial_j E+2\partial_{(i}F_{j)}+t_{ij}, \nn \eea
where $\partial^iF_i=\partial^iV_i=\partial^it_{ij}=t_{ii}=0$.
 The
last two conditions mean that $t_{ij}$ is a transverse and
traceless tensor in three dimensions.  Using this decomposition,
the scalar modes ($A,B,\psi,E$), the vector modes ($V_i,F_i$), and
the tensor modes ($t_{ij}$) decouple completely from each other.
These all amount to 10 degrees of freedom for a symmetric tensor
in four dimensions. Hereafter we consider  tensor modes only.

\subsection{Tensor modes} The field equation for tensor modes is given
by~\cite{My} \be \ddot{t}_{ij}-\frac{\mu^4\kappa^2}{2}
\bigtriangleup t_{ij}
+\frac{\mu^2\kappa^4}{16}\bigtriangleup^2t_{ij}-\frac{\mu\kappa^4\gamma^2}{4w^2}
\epsilon_{ilm}\partial^l\bigtriangleup^2t_j~^m-
\frac{\kappa^4}{4w^4} \bigtriangleup^3 t_{ij}=T_{ij} \ee with
external source $T_{ij}$ and the Laplacian
$\bigtriangleup=\partial_i^2\to-\vec{p}^2$. We could not obtain
 the covariant propagator because of the presence of $\epsilon$-term.
  Assuming a massless graviton propagation along the
$x^3$-direction with $\vec{p}=(0,0,p_3)$, then the $t_{ij}$ can be
expressed in terms of polarization components as~\cite{BS} \be
t_{ij}=\left(%
\begin{array}{ccc}
  t_+ & t_\times & 0 \\
  t_\times & -t_+ & 0 \\
  0 &0  & 0 \\
\end{array}%
\right). \ee Using this parametrization, we find two coupled
equations for different polarizations \bea \ddot{t}_+
-\frac{\mu^4\kappa^2}{2} \bigtriangleup t_+
+\frac{\kappa^4\mu^2}{16}\bigtriangleup^2t_+
+\frac{\kappa^4\mu}{4w^2}\partial_3\bigtriangleup^2t_\times-\frac{\kappa^4}{4w^4}\bigtriangleup^3t_+=T_{+},
\\
\ddot{t}_\times-\frac{\mu^4\kappa^2}{2} \bigtriangleup
t_\times+\frac{\kappa^4\mu^2}{16}\bigtriangleup^2t_\times
-\frac{\kappa^4\mu}{4w^2}\partial_3\bigtriangleup^2t_+-\frac{\kappa^4}{4w^4}\bigtriangleup^3t_\times=T_{\times}.
\eea In order to find two independent components, we introduce the
left-right base defined by \be
h_{L/R}=\frac{1}{\sqrt{2}}\Big(h_+\pm ih_\times\Big) \ee where
$h_L(h_R)$ represent the left (right)-handed modes.  After
Fourier-transformation, we find two decoupled equations \bea
-\omega^2{t}_L+ c^2\vec{p}^2 t_L
+\frac{\kappa^4\mu^2}{16}(\vec{p}^2)^2t_L
-\frac{\kappa^4\mu}{4w^2}p_3(\vec{p}^2)^2t_L+\frac{\kappa^4}{4w^4}(\vec{p}^2)^3t_L=
T_L,
\\
-\omega^2{t}_R +c^2\vec{p}^2 t_R
+\frac{\kappa^4\mu^2}{16}(\vec{p}^2)^2t_R
+\frac{\kappa^4\mu}{4w^2}p_3(\vec{p}^2)^2t_R+\frac{\kappa^4}{4w^4}(\vec{p}^2)^3t_R=
T_R. \eea We have UV-tensor propagators  \be
\label{tenprp}t_{L/R}=-\frac{T_{L/R}}{ \omega^2-c^2\vec{p}^2
-\frac{c^2\kappa^2}{8\mu^2}(\vec{p}^2)^2 \pm
\frac{c^2\kappa^2}{2w^2\mu^3}p_3(\vec{p}^2)^2-\frac{c^2\kappa^2}{2w^4\mu^4}(\vec{p}^2)^3}.\ee
We note that the left-handed mode is not allowed because it may
give rise to ghost
($-\frac{c^2\kappa^2}{2w^2\mu^3}p_3(\vec{p}^2)^2$), while the
right-handed mode is allowed because there is no ghost
($\frac{c^2\kappa^2}{2w^2\mu^3}p_3(\vec{p}^2)^2$). Finally, we
have UV-propagators in the Lorentz-frame with
$p^\mu=(\omega,0,0,p_3)$ as \be t_{L/R}=-\frac{T_{L/R}}{
\omega^2-c^2p_3^2-\frac{c^2\kappa^2}{8\mu^2}p_3^4 \pm
\frac{c^2\kappa^2}{2w^2\mu^3}p_3^5-\frac{c^2\kappa^2}{2w^4\mu^4}p_3^6}.\ee

\subsection{GUP-corrected propagators}
Here we propose that  the GUP-corrected tensor propagators may
take the form \be G_{\rm IR}(\omega,\vec{p})\times{\cal
D}_{D}(\beta\vec{p}^2), \ee where the IR-propagator $G_{\rm
IR}(\omega,\vec{p})$ is defined by \be G_{\rm
IR}(\omega,\vec{p})=\frac{1}{\omega^2-c^2\vec{p}^2}. \ee
 For $D=1$,
its form takes \bea t^{1DGUP}_{ij} &=& -G_{\rm
IR}(\omega,\vec{p})\times{\cal
D}_{1}(\beta\vec{p}^2)~T_{ij}=-\frac{T_{ij}}{(\omega^2-c^2\vec{p}^2)(1+\beta\vec{p}^2)}
\nn \\
=&-&\frac{T_{ij}}{\omega^2-c^2(1-\frac{\beta\omega^2}{c^2})\vec{p}-c^2\beta^2(\vec{p}^2)^2}
\eea which  may be related to  the propagator of $z=2$ HL gravity
defined by the Einstein gravity \be W_{EG}= \mu \int d^3x
\sqrt{g}\Big[R-2\Lambda_W\Big]. \ee
 The $D=2$
GUP-corrected tensor propagator is given by
 \bea
t^{2DGUP}_{ij}&=&-G_{\rm IR}(\omega,\vec{p})\times{\cal
D}_{2}(\beta\vec{p}^2)~T_{ij}=
-\frac{T_{ij}}{(\omega^2-c^2\vec{p}^2)(1+\beta\vec{p}^2)^2}
\nn \\
&=&-\frac{T_{ij}}{\omega^2-c^2(1-\frac{2\beta\omega^2}{c^2})\vec{p}^2-c^2(2\beta
-\frac{\beta^2\omega^2}{c^2})(\vec{p}^2)^2-c^2\beta^2(\vec{p}^2)^3},
\eea where scaling dimensions are given by
$[\beta]=-2,~[\omega]=3,$ and $[c]=2$. This may be related to
UV-tensor propagator (\ref{tenprp}) for $z=3$ HL gravity because
the highest space derivative is sixth order. At this stage, it is
not clear why the $D=2$ GUP-corrected tensor propagator take a
qualitatively  similar form like UV- tensor propagator of $z=3$ HL
gravity except the $p^5$-term. We conjecture that this may be
possible because the $z=3$ HL gravity originates from the detailed
balance condition. Finally, the $D=3$ GUP-corrected tensor
propagator is given by
 \bea
&&t^{3DGUP}_{ij}=-G_{\rm IR}(\omega,\vec{p})\times{\cal
D}_{3}(\beta\vec{p}^2)~T_{ij}=-\frac{T_{ij}}{(\omega^2-c^2\vec{p}^2)(1+\beta\vec{p}^2)^3}
\nn \\
&&=-\frac{T_{ij}}{\omega^2-c^2(1-\frac{3\beta\omega^2}{c^2})\vec{p}^2-3c^2(
\beta-\frac{\beta^2\omega^2}{c^2})(\vec{p}^2)^2-c^2(3\beta^2-\frac{\beta^3
\omega^2}{c^2})(\vec{p}^2)^3-c^2\beta^3(\vec{p}^2)^4 }, \eea which
may be related to the UV-tensor propagator in  $z=4$ HL gravity
because the highest space derivative has eighth order. The  $z=4$
HL gravity was constructed, through the detailed balance
condition, from the new massive gravity~\cite{BHT} \be W_{NMG}=
\int d^3x \sqrt{g}\Bigg[-\mu
(R-2\Lambda_W)+\frac{1}{M}\Big(R_{\mu\nu}R^{\mu\nu}-\frac{3}{8}R^2\Big)\Bigg].
\ee

\section{Discussions}
We have explored a connection between the GUP commutator of
(\ref{crgup}) and the Ho\v{r}ava-Lifshitz gravity (a candidate of
quantum gravity). Explicitly, we have replaced a relativistic
cutoff function $\mathcal{K}(\frac{p^2}{\Lambda^2})$ by a
non-relativistic density function ${\cal D}_{D}(\beta \vec{p}^2)$
to derive GUP-corrected graviton propagators. These were compared
to a UV-tensor graviton propagator in the HL gravity. We point out
that two are qualitatively similar, but the $p^5$-term arisen from
the crossed operation of Cotton and Ricci tensors did not appear
in the GUP-corrected propagators. Also, it is unclear why the
$D=2$ GUP-corrected tensor propagator (not
 the $D=3$ GUP-corrected propagator)
 takes a similar form from the $z=3$ HL gravity. We conjecture
 that it may be related to the detailed balance condition.
Even though our GUP-corrected propagator does not lead to a
precise graviton propagator, this approach will provide a hint to
understand  quantum aspects of  the HL gravity.

A key point  to understand  a connection between two seemingly
different approaches is to recognize  ``effects of quantum
gravity" . The GUP provides naturally a UV cutoff $1/\sqrt{\beta}$
to the LQFT as effects of quantum gravity  through the density
function ${\cal D}_{D}(\beta\vec{p}^2)$. The modified HL gravity
action is composed of higher space derivatives terms from the
detailed balance condition like $R^2,~R^2_{ij},~R_{ij}C_{ij}$ and
$C_{ij}^2$ in addition to $\mu^4R$, to become a power-counting
renormalizable quantum gravity theory.  All these higher
derivative terms modify the tensor propagator into the UV-tensor
propagator in Eq.(\ref{tenprp}) without ghost. We need   a further
study to justify whether there exists an exact connection between
GUP and HL gravity.

Consequently, we have shown that  effects of  quantum gravity are
imprinted on the GUP,  which may explain  the UV-tensor propagator
of the modified HL gravity.

\section*{Acknowledgement}
The author was supported by the SRC Program of the KOSEF through
the Center for Quantum Spacetime (CQUeST) of Sogang University
with grant number R11-2005-021-03001-0.

\end{document}